\documentclass[11pt, preprint2]{aastex}

\newcommand {\belist}{\begin{list}{---}{\setlength{\rightmargin}%
{\leftmargin}}}

\newcounter{arabnum}
\newcommand{\befigcap}{\begin{list}{ {\bf Figure \arabic{arabnum} } %
{ \usecounter{arabnum}} } }
\newcommand{\enfigcap}{\end{list}}

\newcommand{\bequo}{\begin{quotation}}
\newcommand{\enquo}{\end{quotation}}
\newcommand{\bverse}{\begin{verse}}
\newcommand{\everse}{\end{verse}}
\newcommand{\beit}{\begin{itemize}}
\newcommand{\enit}{\end{itemize}}
\newcommand{\been}{\begin{enumerate}}
\newcommand{\enen}{\end{enumerate}}
\newcommand{\ecen}{\end{center}}
\newcommand{\bcen}{\begin{center}}

\newcommand{\begeq}{\begin{equation}}
\newcommand{\eneq}{\end{equation}}
\newcommand{\befig}{\begin{figure}}
\newcommand{\enfig}{\end{figure}}
\newcommand{\ferrh}{\mbox{$\rm{[Fe/H]\:}$}}

\newcommand{\kmsec}{\mbox{${\rm \: km\:s^{-1}}$}}

\newcommand{\msol}{\mbox{$\: M_\odot$}}

\newcommand{\uvby}{\mbox{\it uvby }}

\slugcomment{}

\shorttitle{Dusty Pre-Mainsequence F Stars}
\shortauthors{Suchkov et al.}

\begin{document}

\title{Candidate Pre-Mainsequence F~Stars with Circumstellar Dust
Identified Using Combined 2MASS 
and \uvby Data}
\author{A. A. Suchkov\altaffilmark{1}\altaffiltext{1}{Space Telescope Science
Institute, operated by AURA Inc., under contract with NASA,
3700 San Martin Dr., Baltimore, MD 21218; suchkov@stsci.edu.}, 
A. B. Schultz\altaffilmark{1,2}\altaffiltext{3}{Science Programs,
Computer Sciences Corporation, 3700 San Martin Dr.,
 Baltimore, MD 21218; schultz@stsci.edu.}, and
C. M. Lisse\altaffilmark{3}\altaffiltext{3}{University of Maryland,
Dept. of Astronomy, College Park, MD 26792; lisse@astro.umd.edu}
}

\begin{abstract}

We propose a method that uses near-infrared plus \uvby photometry 
to identify  potentially extensive circumstellar dusty environment
about  F~and~A stars.  The method  has been applied to a sample of
$\sim 900$ metal rich reddened F~stars with 2MASS and {\it uvby} data,
suggesting the presence of circumstellar dust emitting in the
near infrared for $\sim$~70 stars.
The $\log T_e - M_V$ diagram suggests that most, if not all, of them are
likely pre-mainsequence (PMS). They seem to be consistent with being a
continuation of the class of Herbig~Ae/Be PMS stars into the spectral
type F.  Their number drops sharply downward of $\log T_e \sim 3.84$
(spectral types later than $\sim$~F5), which may provide new clues to
the PMS evolution of $1-2\msol$ stars.  We present a list of 21
most conspicuous candidate stars with circumstellar dust.  About half
of them are associated with the extended star-forming region around
$\rho$~Oph. The brightest of these 21 stars, with $V \lesssim 7.5$,
turn out to be IRAS sources, suggesting the presence of heated dust
emitting in the far infrared. Also in this list, HD 81270 is reported 
as a very unusual star moving away from the Galactic plane at a 
projected speed of 70 \kmsec.  

\end{abstract}


\section{Introduction}

The detection of circumstellar dust disks provides us with an
unprecedented opportunity to study the formation and evolution of
exo-solar systems (see, e.g., Beckwith 1999) as well as  the
PMS stellar evolution (e.g., van den Ancker et al. 1997),
given that many of these systems are associated with PMS stars.  
The bulk of the known PMS stars
are comprised by Herbig Ae and T~Tauri stars.  The presence of large
amounts of dust about many of these stars has been inferred from IRAS
observations (Aumann et al. 1984; Aumann 1985; Gillett 1986; Backman
\&\ Gillett 1987; Walker \&\ Wolstencroft 1988; Backman \&\ Paresce
1993). The IRAS survey still remains a major source to search for candidate
dust-disk systems (e.g., Mannings \&\ Barlow 1998). Some
400 IRAS candidates exhibit an unusually large far-infrared (FIR) excess
(e.g., Stencel \&\ Backman 1991, Mannings \&\ Barlow 1998), interpreted
as being due to thermal emission of hot circumstellar dust ($30 < T <
1000$~K).
In contrast to IRAS systems whose large FIR excess results
from a high-density concentration of warm dust, there are young stellar
objects (YSO) that show relatively small FIR excess. They include
T~Tauri stars exhibiting  spectral energy distribution (SED) that
declines longward of $\sim 2$~\micron\  (YSOs with SED of Class II and
Class III); these stars are interpreted as advanced evolutionary phases
of T~Tauri systems, with only a fraction of the original dust mass
remaining in the disk (e.g., Lada and Wilking 1984; Andr\'{e} \&
Montmerle 1994).  Less massive dust disks or envelopes may not be
detectable in the far infrared.  Disk clearing may have
started for these stars, but it may still be possible to detect the
disk remnant material in the near infrared through reflected light.

In this paper we present a new criterion that uses a combined set of
2MASS and {\it uvby} data to identify  F~stars with circumstellar
dust.  The candidate stars found with this method appear to possess
dusty disks or envelopes radiating the reflected and/or thermal
near-infrared (NIR), causing a star to look in the NIR brighter than
expected for a reddened normal F~star at the same visual magnitude
$V$.  Many of them are {\it not} IRAS sources and may therefore
represent a dusty environment substantially different from that found
among the IRAS sources.

\section{Data}
Starting with a sample described in Suchkov \&\ McMaster
(1999), which includes F~stars  with {\it uvby} colors from 
Hauck \&\ Mermilliod (1998) and observed by the {\it Hipparcos}, we
found ${\sim 5230}$ counterparts in the 2MASS Point Source Catalog.
Some of the stars among those with high reddening are, in fact,
of the spectral type A rather than F. The reason is that the selection
for the original sample was made based on the $(b-y)$ color,
specifically, $0.222 < b-y < 0.390$. For unreddened stars, the latter
criterion effectively selects stars later than spectral type A.
However, if reddening is present, stars with intrinsically bluer
colors, including A~stars, can also be picked up by this criterion.  The
overall fraction of A~stars in our sample is very small, and  for the
sake of simplicity, we'll be referring to the sample stars as F~stars.
The criterion also selects a small fraction of early~type G~stars.

\begin{figure}
\plotone{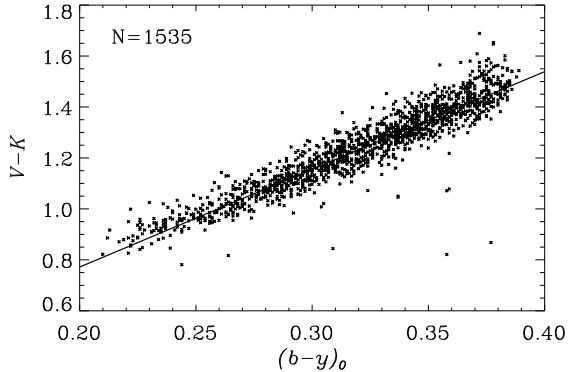}
\caption
{ $(V-K)$ versus ${(b-y)_0}$ diagram for F~stars within 80~pc. 
The thick line is the linear regression used to compute $\delta (V-K)_{b-y}$. 
}
\label{f1}
\end{figure}

The \uvby data were used to derive reddening $E_{b-y}$, effective 
temperature, 
and metallicity as described in Suchkov \&\ McMaster (1999) and Suchkov (2001), 
based on the algorithm in Moon (1985) as well as the Schuster \&\ Nissen (1989) 
metallicity calibration. Extinction was  estimated as 
\begeq
	A_V = \frac{E_{b-y}} {0.74}\, (3.354 + \frac{0.0406 E_{b-y}} {0.74}). 
\label{eq-av}
\eneq
This equation follows  from the results in Strai\v{z}is (1977), Crawford 
(1975), and Sud\v{z}ius (1974). 
Absolute magnitude was computed from the $V$ magnitude corrected for 
extinction, using the {\it Hipparcos}  parallax; $V$ magnitudes were taken from 
the {\it Hipparcos}  catalog.

\section{ Circumstellar dust from intrinsic near-infrared excess } 

The  sample stars exhibit a tight temperature-induced correlation between the 
temperature-sensitive \uvby color index $(b-y)_0$ and $V-K$.
In Figure~\ref{f1}, this correlation is shown for stars  within 80~pc
with small or no reddening, $E_{b-y} < 0.015$. The scatter about the
regression line, $ \sim 0.05$ after 3-sigma correction, 
is due to both photometric
errors and real physical effects,  mainly related to absolute magnitude,
duplicity, and remaining reddening. It reduces to $\sigma_{V-K} \sim 0.035$
if the range for these parameters is narrowed down to $2 < M_V < 3$,
$E_{b-y} < -0.005$, and binaries are excluded. This should be close 
to the value resulting from photometric errors only.   
The linear regression shown in Figure~\ref{f1} remains, however,
the same, within uncertainties, as for the constrained sample. 
It was used to compute the deviation from the regression line for all 
sample stars, 
\begeq
	\delta (V-K)_{b-y} = (V-K) - (V-K)_{b-y}. 
\label{eq-vkby0}
\eneq
A star having an excessive NIR flux with respect to the one expected for 
an unreddened standard  star at the  same effective temperature would 
deviate upward from this line,
\begeq
\delta (V-K)_{b-y} > 0. 
\label{eq-excess}
\eneq

\begin{figure}
\plotone{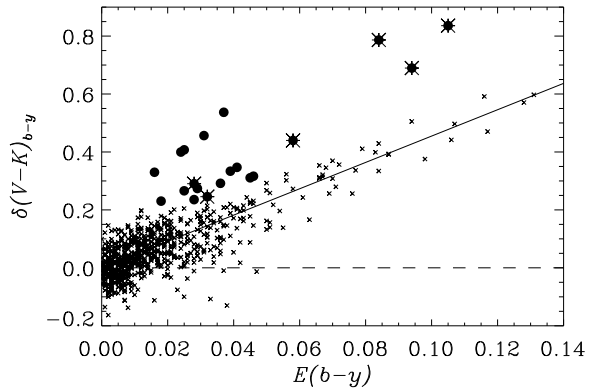}
\caption
{Near-infrared excess  versus reddening  for metal-rich 
(\ferrh~$ > 0$) F~stars. The most prominent candidate dust-disk stars 
are shown as filled circles, with IRAS sources among them indicated by 
overlaid asterisk.
The solid line is the relationship between $\delta(V-K)_{b-y}$ and
$E_{b-y}$ expected from the reddening effect only (see text for details).
\label{f2}
}
\end{figure}

Circumstellar dust can contribute to observed infrared excess
in two quite different ways. First, this material would absorb stellar
light stronger at shorter wavelengths, which adds up  with the effect
of interstellar extinction and results in flux deficiency in the blue
and visual with respect to the flux in the infrared (the reddening
effect).  The part of the infrared excess due to this effect will be
referred to as {\it apparent} infrared excess.  Second, a dust disk is
a source of reflected and thermal radiation.  Unlike circumstellar and
interstellar extinction, this radiation actually adds to the obscured
stellar flux  in the infrared. The part of the near-infrared excess due
to this extra flux will be referred to as {\it intrinsic} near-infrared
excess.  Isolating  the latter part in the observed total infrared
excess would thus amount to the detection of circumstellar dust.
\begin{figure}
\plotone{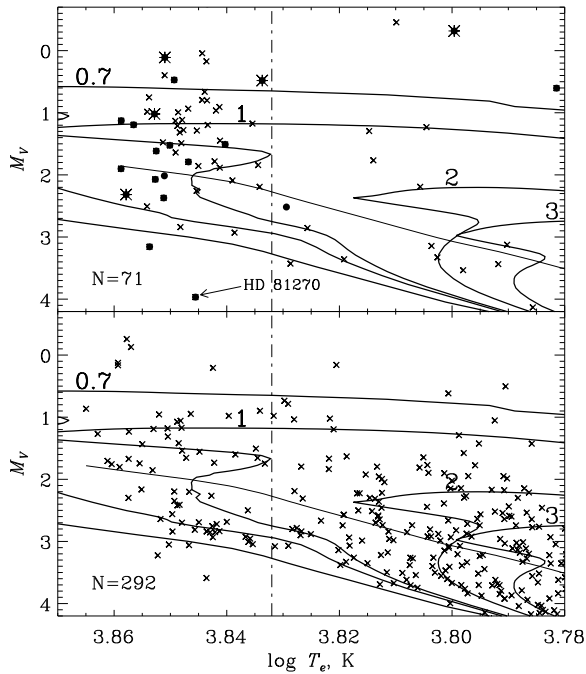}
\caption
{Comparison of  stars with and without intrinsic near-infrared excess,
$\Delta(V-K)_{b-y} > 0.07$ (upper panel) and $ < -0.03$ (lower panel), 
respectively.  
Overlaid are isochrones for the Hyades chemical composition 
(Chaboyer et al. 1999) labeled by age in  Gyr.
\label{f3}
}
\end{figure}

The two parts of the  infrared excess can be separated as follows.
The reddening effect in the total near-infrared excess as  defined by 
equations~(\ref{eq-vkby0}) and (\ref{eq-excess}) is clearly  seen in  
the  $E_{b-y}-\delta (V-K)_{b-y}$ diagram in Figure~\ref{f2}. 
In the diagram, the bulk of the reddened stars  tightly envelope 
the line representing the relationship between $\delta(V-K)_{b-y}$
and $E_{b-y}$ expected from reddening only,
\begeq
\delta(V-K)_{b-y} = A_V (E_{b-y}),
\label{eq-line}
\eneq
where $A_V $ is defined by equation~(\ref{eq-av}), and extinction in $K$ is 
neglected in comparison with $A_V$.  Suppose a star has a circumstellar 
dust disk reradiating stellar optical and UV light in the infrared as well as 
reflecting unabsorbed stellar infrared radiation.
In the $E_{b-y} - \delta(V-K)_{b-y}$ diagram, such a star would 
deviate upward from the line given by equation~(\ref{eq-line}), the amount
of deviation representing a measure of  intrinsic infrared excess. 
Therefore, we can use this deviation as a criterion identifying candidate 
dusty stars:
\begeq
\Delta(V-K)_{b-y} \equiv  \delta(V-K)_{b-y} - A_V (E_{b-y}) > 0. 
\label{eq-ddelta-def}
\eneq
We have applied this criterion to our sample of F~stars. Given
that large masses of dust are expected only in very young systems, it
makes sense to start a search for  dusty stars by examining  metal rich
stars because high metallicity typically implies a young age.
Therefore, we have restricted our search sample to stars 
with [Fe/H]~$ > 0 $. Also, since we expect a dusty system to be
reddened, only stars with $E_{b-y} > 0$ have been considered.  These
constraints reduce the working sample to $\sim 900$ stars. These stars are
shown in Figure~\ref{f2}; those above  the  reddening line
are predicted, in accordance with our criterion, to possess 
circumstellar dust.

The most conspicuous candidate dusty stars are shown in 
Figures~\ref{f2}~and~\ref{f3} as filled circles; they are also listed in
Table~\ref{tbl-selected}.  These 21 stars have $E_{b-y} > 0.01$ and
$\Delta (V-K)_{b-y} > 0.1$.  About half of them are spatially
associated with the giant molecular complex centered at the site of
ongoing star formation in $\rho$~Oph.

IRAS fluxes in Table~\ref{tbl-selected}, $F(12 \; \mu)$, $F(25 \;
\mu)$, and  $F(60 \; \mu)$, are from the IRAS Faint Source Catalog,
Version 2.  In total, we found seven out of the 21 selected candidate
PMS dusty F~stars to be IRAS sources (marked in Figure~\ref{f2} by
overlaid asterisk), which suggests the presence of heated circumstellar
dust about them. These 7 stars are brighter than $V \sim 7.5$~mag (with
a notable exception for HD 144432, which is 8.188 in $V$),  which may
hint that the thermal emission from possible dust disks around fainter
F~stars is simply below the detection limit of the IRAS; however,
an intrinsic difference between the IRAS and non-IRAS stars
cannot be ruled out.  We have also searched through the ISO
archive (using SIMBAD interface) to see if any of the stars in
Table~\ref{tbl-selected} was among the ISO pointed observations, but
found none.

To gain insight into the nature of the predicted stars with
circumstellar dust, we have examined the $\log T_e - M_V$  diagrams for
more extreme, 2~$\sigma$ candidates, $\Delta(V-K)_{(b-y)_0} > 0.07$, and compared
it with the one for ``noncandidates'' defined by $\Delta(V-K)_{(b-y)_0} <
- 0.03$, selecting both groups from the sample shown in
Figure~\ref{f2}.  Inspection of Figure~\ref{f3} reveals a large
difference between the two diagrams. While the stars in the lower panel
scatter across the entire temperature range displayed in the diagram,
the stars in the upper panel exhibit a heavy concentration at the hot
end of the temperature range where only very young stars can reside.
Thus, the stars predicted to possess dusty disks (upper panel), are
mostly very young whereas the non-candidates (lower panel) are
dominated by a much older, on average, population. This is exactly what
one would expect comparing stars with and without circumstellar dust.

\begin{figure}
\plotone{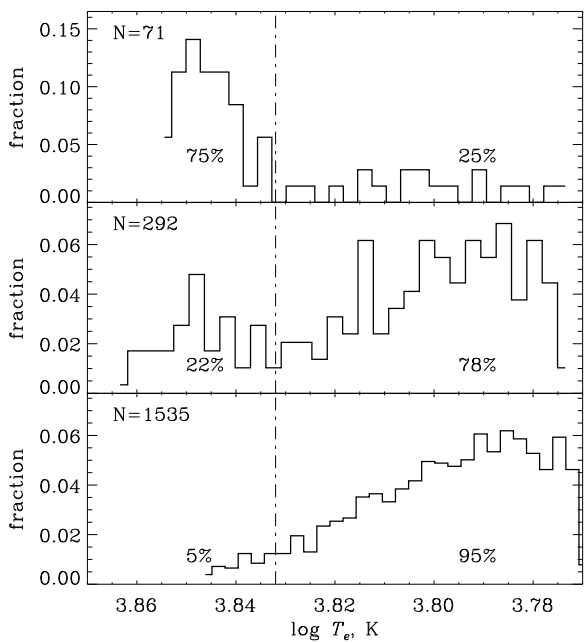}
\caption
{Temperature distribution of the stars shown in Figure~\ref{f3}
(upper and middle panels) and Figure~\ref{f1} (lower panel).  
Percentage of stars to the right and left of the vertical line 
is indicated.
\label{f4}
}
\end{figure}

The prominent clustering of the stars in the upper panel in
Figure~\ref{f3} within a narrow temperature range of $3.86 <
\log T_e < 3.83 $ (see also Figure~\ref{f4}), which span about
3~mag in $M_V$, offers a further clue to the nature of these objects.
The large spread of absolute magnitude they exhibit is obviously
inconsistent  with the effects of stellar evolution away from the main
sequence; if it were, the dominant population in the $\log T_e - M_V$
diagram would have been at cooler rather than hotter temperatures,
simply because the initial mass function increases toward lower
masses.  To put it differently, once we ask the  question: ``Where are
stars in that diagram that must be numerous downward of $\log T_e \sim
3.83$?'', we would immediately realize that most of the hot stars above the
ZAMS in Figure~\ref{f3} are {\it not} post-ZAMS stars but rather are 
at various phases of PMS evolution---which  
perfectly  matches the inferred  presence of  dust around these stars.

One may notice that also in the lower panel in Figure~\ref{f3}
and the respective middle panel in Figure~\ref{f4}, there are many 
hot stars separated from the
bulk of cool stars by a poorly populated temperature interval around
$\log T_e \sim  3.832$.  The same arguments as above make us believe that
these are also mostly PMS stars. Additional argument is that their 
reddening is about twice as large as that of the cool stars, 
$E(b-y) = 0.028 \pm 0.003$ versus $ 0.012 \pm 0.001$; this should be 
expected if a  significant part of it, if not all, occurs in a dense
circumstellar environment of a PMS star. The absence of excessive NIR 
emission apparently indicates that the dust is colder than in the 
PMS candidates in the upper panel in Figures~\ref{f3} and \ref{f4}.  
Incidentally, these latter candidates are less reddened, 
$E(b-y) = 0.021 \pm 0.001$, then the stars with presumably cold dust, 
which should be expected if warm dust becomes more dominant 
when much of the original dust content has been dispersed.

Note that a  large fraction of dusty PMS candidates inferred for the sample
shown in Figure~\ref{f3} is not at all typical for the general
population of the field F~stars. The temperature distribution
of a sample that is more or less representative of a general population 
(the sample shown in Figure~\ref{f1}) is displayed in the lower panel 
of Figure~\ref{f4}. It is obvious that for this sample the fraction of 
stars in the temperature range dominated by PMS candidates is quite low
and consistent with the expectations for an IMF declining toward higher
stellar masses.  

The dusty PMS candidates are contiguous in the $\log
T_e - M_V$  diagram with A stars that are known to contain a population
of PMS emission-line, dusty stars usually
referred to as a group as  Herbig Ae/Be (HAeBe) stars.  The properties
of the candidates and the fact that they are adjacent to A stars
suggest that they represent a continuation of the population of HAeBe
stars into the spectral class F.  Th\'{e} et al. (1994) extended the
criteria for membership of the class of HAeBe stars to include emission
line objects of spectral types earlier than F8. However, the number of
stars later than  F2 in  their catalog of HAeBe objects is very small,
only about half a dozen. The candidate PMS dusty F~stars identified
in this paper would represent an  enormous increase in the  number of F
stars later than F2 that appear to be akin to HAeBe stars. 

As discussed above, the stars in the upper panel of
Figure~\ref{f3}  are concentrated within a narrow temperature
range, $3.860 < \log T_e < 3.832$ (spectral range $\sim$~F2~to~F5),
their number plummeting downward of $\log T_e \sim 3.84$.  This may
have interesting implications pertinent to the nature of the class of
HAeBe objects extended into the  spectral range F, suggesting that such
objects get extremely rare redward of F5.  This phenomenon is likely
related to fundamental properties of the PMS evolution of $1-2\msol$
stars. So the detection of the  cut-off in the
temperature (spectral type) distribution of HAeBe+F objects at $\log
T_e \sim 3.84$ (Sp~$\sim$~F5) should be expected to provide new
insight into the physics of PMS stars.

In summary, we have proposed a new criterion to identify F~stars with
warm circumstellar dust emitting in the near infrared. Application of
this criterion to a sample of $\sim 900$ metal rich, moderately
reddened  F~stars resulted in detection of $\sim 70$ 2-sigma candidates.
They appear to be PMS stars, representing a continuation of the class of 
Herbig Ae/Be stars into the spectral range F and exhibiting a cutoff 
redward of $\sim$~F5. There is also evidence for a population of
PMS stars with cold dust undetectable in the near infrared. 



\begin{deluxetable}{lccccccccc}  
\tabletypesize{\scriptsize}
\tablewidth{0pt}
\tablecaption{ Photometric properties of selected stars with intrinsic
infrared excess
 \label{tbl-selected} 
}
\tablehead{
\colhead{Star} & \colhead{Sp} & \colhead{V} & \colhead{J}& \colhead{H}& \colhead{K}& \colhead{V-K} & \colhead{F(12 $\mu$)} & \colhead{F(25 $\mu$)} & \colhead{F(60 $\mu$)} \\
  \colhead{}  
& \colhead{}  
& \colhead{}  
& \colhead{}  
& \colhead{}  
& \colhead{}  
& \colhead{}  
& \colhead{(Jy)}  
& \colhead{(Jy)}  
& \colhead{(Jy)}  
}
\startdata
HD18466 & A2/A3V & 6.286 & - & - & 4.678 & 1.608 & 0.555  & 0.113  & 0.112  \\

HD71766 & F2III & 5.995 & - & - & 4.678 & 1.317 & 0.495  & 0.105  & 0.113  \\

HD73378 & A9III & 8.000 & 7.178 & 6.997 & 6.948 & 1.052 & - & - & - \\

HD81270\tablenotemark{a} 
& F6V   & 8.950 & 7.915 & 7.636 & 7.553 & 1.397 & - & - & - \\

HD88021 & F5 & 6.678 & 5.405 & - & 4.915 & 1.763 & 0.455  & 0.231  & 0.133  \\

HD115984 & F3V & 8.191 & 7.405 & 7.229 & 7.121 & 1.070 & - & - & - \\

HD120850 & F0V & 7.362 & 6.605 & 6.449 & 6.341 & 1.021 & - & - & - \\

HD121036 & F2V & 9.549 & 8.609 & 8.458 & 8.370 & 1.179 & - & - & - \\

HD133411 & F0V & 7.911 & 7.014 & 6.888 & 6.831 & 1.080 & - & - & - \\

HD144432\tablenotemark{b}
& A9/F0V & 8.188 & 7.115 & 6.541 & 5.894 & 2.294 & 7.62  & 9.23  & 5.68  \\

HD151070 & F5III & 6.871 & 5.612 & 5.297 & 5.190 & 1.681 & 0.389  & 0.116  & 0.131  \\

HD152500 & F0V & 8.202 & 7.343 & 7.227 & 7.125 & 1.077 & - & - & - \\

HD153777 & F2IV & 8.435 & 7.547 & 7.331 & 7.224 & 1.211 & - & - & -  \\

HD156441 & A9V & 9.100 & 8.410 & 8.245 & 7.930 & 1.170 & - & - & - \\

HD158063 & F8 & 7.383 & 6.088 & 5.745 & 5.666 & 1.717 & 0.219  & 0.044  & 0.092  \\

HD160410 & F0V & 8.210 & 7.416 & 7.109 & 6.921 & 1.289 & - & - & - \\

HD164200 & F3V & 8.410 & 7.445 & 7.306 & 7.173 & 1.237 & - & - & - \\

HD172550 & F0V & 8.922 & 8.116 & 7.982 & 7.862 & 1.060 & - & - & - \\

HD195678 & A9V & 9.133 & 8.297 & 8.127 & 8.077 & 1.056 & - & - & - \\

HD212500 & F4III & 6.950 & 6.009 & 5.863 & 5.791 & 1.159 & - & - & - \\ 

HD222355 & F5 & 7.548 & 6.219 & 5.907 & 5.751 & 1.797 & 0.204  & 0.186  & 0.173  \\


\enddata
\tablenotetext{a}{Flagged in SIMBAD as high proper-motion. 
With tangential velocity of 101 \kmsec, the star moves away from the Galactic 
plane at an extremely high velocity of 70~\kmsec. 
The star is very unusual in other respects; it is extremely 
underluminous (below the Hyades ZAMS by $\sim 1.2$~mag), shows a signature
of fast rotation and/or anomalously low surface gravity, and its metallicity
exceeds the mean metallicity of the Hyades F~stars by $\sim 0.2$~dex---
all suggesting that something dramatic happened to this star recently.
}
\tablenotetext{b}{Flagged in SIMBAD as pre-mainsequence. Listed as a candidate
disk-bearing system in Mannings \&\ Barlow (1998) on the basis of its
IRAS emission.} 
\end{deluxetable}


\end{document}